\documentstyle[aps,preprint]{revtex}
\newcommand{\gs}{\hbox{\kern1.5mm\raise0.8mm\hbox{$>$}\kern-3.2mm
                 \raise-1.0mm\hbox{$\sim$}\kern1.5mm}}
\newcommand{\ls}{\hbox{\kern1.5mm\raise0.8mm\hbox{$<$}\kern-3.2mm
                 \raise-1.0mm\hbox{$\sim$}\kern1.5mm}}
\newcommand{\ts}{\left(\frac{\tau}{{\rm sec}}\right)}
\newcommand{\mY}{\left(\frac{mY}{{\rm MeV}}\right)}
\newcommand{\mYi}{\left(\frac{{\rm MeV}}{mY}\right)}
\begin{document}
%\tighten

{\hbox to\hsize{May 1994 \hfill SNUTP 94-48}}
{\hbox to\hsize{hep-ph/9405385}}

\bigskip

\bigskip

\title{Dark Matter and Structure Formation\\
with Late Decaying Particles}
\author{Hang Bae Kim and Jihn E. Kim}
\address{Center for Theoretical Physics and Department of Physics\\
         Seoul National University\\
         Seoul 151-742, Korea}
\maketitle
\begin{abstract}
Since it became evident that the CDM model for cosmic structure formation
predicts smaller power on large scales than observed, many alternatives have
been suggested.  Among them, the existence of late decaying particle can cure
it by delaying the beginning of the matter domination and increasing the
horizon length at that time.  We discuss the realization of this scenario and
present the light neutrino and the light axino as possible examples of working
particle physics model.  We point out that the increased power at sub-galaxy
scale predicted by this scenario could lead to rich sub-galaxy structures.
\end{abstract}
%\pacs{98.80.Cq, 95.35.+d, 14.80.Ly}

\newpage

\section{Introduction}

In the past decade, the cold dark matter (CDM) model has attracted a great
attention as a possible theory of the cosmic structure formation.  It is
characterized by a spatially flat universe with $\sim$90\% of the mass density
formed by cold dark matter and the scale invariant primordial fluctuation
spectrum.  It was supported by its successful features, the simplicity of its
underlying assumptions and its link to the physics of the early universe.  The
basic idea that the large scale structure seen today evolved from very small
primordial density fluctuations was strengthened by the recent detection of
large scale anisotropies in the cosmic microwave background \cite{s92}.  The
candidates for cold dark matter were provided by particle physics models.  The
scale invariant fluctuation spectrum was naturally obtained from inflation
which is an indispensable ingredient of current cosmology.

Recently, however, it became evident that this model is in conflict with
observations \cite{defw92}.  The main difficulty was that the predicted power
spectrum could not fit well with all the observational data simultaneously.
The reason might be attributed to the relative smallness of the horizon length
when the universe passes from relativistic matter domination to nonrelativistic
matter domination which is the unique length scale characterizing the simple
CDM model.  Therefore, a more complicated model may be needed.

There are two main variables in the structure formation theory, the content of
matter in the universe and the shape of the primordial density fluctuations.
A variation of two ingredients leads to a different prediction on the power
spectrum.  Actually several variants yield better agreements with observations
than the cold dark matter model \cite{bbe87}. The cold plus hot dark matter
model attracted broad interest recently in this regard \cite{dss92}.  The warm
dark matter model is another appealing possibility which draws new attention
\cite{dw94}.  Altering the primordial fluctuation spectrum is another
alternative \cite{sbb89}.  In addition, there is a way to maintain the matter
content as cold dark matter by increasing the horizon length at the time of
radiation matter equality by appropriate amount.  Introduction of cosmological
constant is one way to achieve this scenario \cite{esm90}.  The existence of
late decaying particles which shift the beginning of matter domination era is
another possibility \cite{bbe87}.  At present, the large scale structure
phenomenology does not favor any specific one among these variants.

{}From the point of view of particle physicists, we have a similar situation
because most dark matter candidates have no strong support for their existence
from the particle physics phenomenology and variant models require unconfirmed
new physics.  Therefore, the present criterion in particle physics is the
necessity and plausibility of the models accommodating these variants.  In view
of this criterion, the problem in the hot plus cold dark matter model is the
one of explaining the comparable amount of hot and cold dark matter without
attributing it to a mere chance.  At first glance, the massive neutrino and the
lightest supersymmetric particle as the hot and cold dark matter candidates
seem to have no relation between them.  Recently, some suggestions were made,
connecting the amount of cold dark matter to that of hot dark matter which is
usually assumed to be neutrinos \cite{kms93}.  For the cold plus hot dark
matter models, one usually introduces additional parameters which need
explanation.  In a similar manner, introduction of late decaying particles
faces the problem of explaining the simultaneous existence of cold dark matter
with critical density and a late decaying particle with appropriate lifetime.
It would face the similar problem of plausibility as the hot plus cold dark
matter model does.  At present, however, it is welcome to have a working model
for it.  This is our philosophy in this paper.  We pursue for the possibility
of late decaying particles dominating the energy density of the universe for
some time after nucleosynthesis and decaying to extremely relativistic, weakly
interacting particles.  They shift the time of radiation matter equality, which
leads to the elongation of the horizon length at that time and the suppression
of the power on large scales compared to the value predicted in the simple CDM
model.  This idea was first pointed out in Ref.~\cite{bbe87}.  Later, the
17 keV neutrino was used to implement the idea \cite{be91}, but its existence
is questioned now \cite{n92}.  Recently, this idea has been rediscovered
in the light axino decay \cite{ckk94}, through the naive expectation that the
late decaying particles producing relativistic particles can mimick the cold
plus hot dark matter scenario.

In this paper, we emphasize the role of late decaying particles in cosmic
structure formation and obtain the conditions on the relic amount and lifetime
applicable to the generic case.  Then the explicit models for late decaying
particles are presented.  The plan of the paper is as follows.  In section II,
we give a brief review on the structure formation theories and compare the
several alternatives to the simple CDM model.  In section III, the role of
late decaying particles in structure formation is explained and the condition
for better agreement with the observational data is presented.  In section IV,
we present the scenarios for the light neutrino and the light axino as
explicit models for the cosmology of late decaying particles.  Section V is a
discussion and a conclusion.

\section{The Structure Formation Theories and Dark Matter}

Structure formation and dark matter are closely related.  Any theory for the
formation of cosmic structure must specify two chief ingredients: the amount
and nature of the material that fills the universe, and the properties of the
seed density fluctuations from which galaxies and clusters developed.

The amount of each component of materials in the universe is represented by
the ratio of its mass density to the critical density:
$\Omega_i=\rho_i/\rho_{\rm crit}$.  There are two widely accepted beliefs
concerning the amount of matter \cite{kt90}.  One is that the total amount of
matter is nearly critical ($\Omega\equiv\sum_i\Omega_i=1$).  The other is that
the amount of baryons is less than 10\% of the critical density
($\Omega_B \ls 0.1$).  The former follows from most inflationary scenarios.
The strongest argument for the latter is the light element production in
nucleosynthesis.  The gap between $\Omega$ and $\Omega_B$ can be filled by the
existence of nonbaryonic dark matter or/and the cosmological constant.  Dark
matters are classified into hot, warm and cold dark matters by the
characteristic scales they develop during the evolution of seed density
fluctuations.

At present, two types of origins for the seed density fluctuations are widely
discussed: inflation and topological defects.  In the inflationary scenario
the quantum fluctuations of the inflaton field are changed into the density
fluctuations, which is featured by the scale invariant fluctuation spectrum
\cite{gp82}.  The topological defects are created during the cosmic phase
transitions in some models with a spontaneously broken global symmetry.
Cosmic strings and textures have been discussed in this context.

The seed density fluctuations evolve under the act of gravity.  For small
fluctuations, the evolution is studied by the numerical integration of coupled
perturbed-Einstein-Boltzmann equations.  With the seed density fluctuation
spectrum as initial condition, we obtain the evolved linear fluctuation
spectrum.  For large fluctuations, numerical simulation is used to show the
nonlinear dynamics of clustering.  Much work has been done with various dark
matter contents and seed density fluctuation spectra.  In the literature, one
can find the numerical fit formulae for the evolved linear fluctuation spectra
and the results of numerical simulations for many cases \cite{p82}.  We can
understand the qualitative features of these spectra by the characteristic
scales inherent to the seed density fluctuations or developed by the matter
content during the evolution.  Important ones are the following.  The Jeans
length $\lambda_J$ separates the gravitationally stable and unstable modes.
The free streaming length $\lambda_{FS}$ is a scale below which fluctuations
in a nearly collisionless component are damped due to free streaming.  Another
significant length scale is the horizon length at the time of the radiation
matter equality $\lambda_{EQ}$.  This scale appears because the growth of the
density fluctuation of nonrelativistic matter within the horizon is suppressed
during the radiation dominated era while begins as the matter domination era
starts.

Among the various models, the CDM model was most successful in the past decade.
The CDM model assumes a flat ($\Omega=1$) universe in which $5\sim10$\% of the
critical density is provided by ordinary matter and the rest by the weakly
interacting, yet unknown, cold dark matter.  In addition, the seed fluctuations
are assumed to be of the scale invariant form.  The key assumptions of this
model fit nicely with ideas on inflation.  The evolved fluctuation spectrum is
given by \cite{p82}
\begin{equation}
|\delta_k|^2 = \frac{Ak}{(1+\alpha k+\beta k^{3/2}+\gamma k^2)^2}
\label{spectrum}
\end{equation}
where $A$ is a normalization constant and $\alpha=1.7\ l$,
$\beta=9.0\ l^{3/2}$, $\gamma=1.0\ l^2$, and
$l=(\Omega h^2)^{-1}\theta^{1/2}$Mpc.
The scaling of $l$ with $(\Omega h^2)^{-1}\theta^{1/2}$ simply reflects the
horizon length at the time of radiation matter equality which is given by
\begin{equation}
\lambda_{EQ} \simeq 30(\Omega h^2)^{-1}\theta^{1/2} {\rm\ Mpc},
\end{equation}
where $\theta=\rho_{\rm rel}/1.68\rho_\gamma$ measures the present energy
density of extremely relativistic particles and is 1 for photon and three
massless neutrino flavors.  It is independent of the nature of CDM and
$\lambda_{EQ}$ is a single physical length scale characterizing the above
spectrum.  The power spectrum of the model is compared to that extracted from
the observational data \cite{vpgh92} in Fig.~1.  It shows that with the
standard choice $\Omega h=0.5$ the model does not fit the data.  With the
normalization fixed at COBE data, we can say that the standard CDM model
predicts more power at small scales than observed.  From the figure, we
intuitively reason that the problem in the CDM model is that it has a
relatively small $\lambda_{EQ}$.

Several cures for the excessive power of the simple CDM model at small scales
have been considered.  We put them into four categories:
({\it i\/})   assuming large biasing or antibiasing,
({\it ii\/})  the initial fluctuation spectrum with less power at small scales,
({\it iii\/}) the mixed dark matter contents, and
({\it iv\/})  increasing $\lambda_{EQ}$.
Biasing is the modulation of the galaxy distribution by processes associated
with galaxy formation. Though a complicated version in this category met some
successes, we don't know its detailed physical mechanism.
The initial fluctuation spectrum with less power at small scales can be
obtained in fine tuned inflationary models \cite{sbb89} or in cosmic string
and texture models.  But they require peculiar models with fine tuning of
parameters.
The mixed dark matter content will change the evolution of the initial power
spectrum.  Cold plus hot dark matter model was shown to be successful in this
regard.  With the massive neutrino as hot dark matter, the best fit was
obtained for $\Omega_{\rm CDM}=0.6$, $\Omega_{\nu}=0.3$ and
$\Omega_{\rm B}=0.1$ \cite{w92}.  While this model met many cosmological
successes, it also has a plausibility problem: how to explain the comparable
amount of hot and cold dark matter without attributing it to a mere chance.
It might lead to fine tuning again.  Warm dark matter model may give a
promising result.
In increasing $\lambda_{EQ}$, the simplest idea is to lower $\Omega$ or $h$ to
make $\Omega h\simeq0.2\!-\!0.3$. (See Fig.~1.  We allow small biasing.)  But
this conflicts with other observational data \cite{kt90}.  Introduction of
small cosmological constant allows $\Omega_{\rm CDM}=0.2$ in the flat universe,
resulting in the increase of $\lambda_{EQ}$ \cite{esm90}.  But this requires
severe fine tuning of the cosmological constant, which is not solved yet in
particle physics.  Late decaying particles can lead to the retardation of
matter domination with the desired increase in $\lambda_{EQ}$.  We will
discuss this scenario in detail in the following section.

\section{The Role of Late Decaying Relic Particles in Structure Formation}

In the CDM model, the power on small scales can be reduced by delaying the
beginning of matter domination era.  Since the density fluctuations of cold
dark matter grow significantly during the matter domination era, the growing
time for small scales is reduced as much as one delays the time for the small
scales to enter the horizon before the radiation matter equality.  The
$\Omega h=0.2$ CDM model is very successful in this way.  But the problem is
that one wants to get the same success while maintaining $\Omega=1$.  One way
to achieve this is by postulating a late decaying relic particle which
dominates the energy density of the universe for some time and decays into
extremely relativistic particles.  By producing relativistic particles, the
universe enters into a radiation dominated era around the time of its decay.

Consider a late decaying relic particle $X$ with the mass $m$, the lifetime
$\tau\gg1$ sec (the time of neutrino decoupling) and the ratio of the number
density to entropy density $Y$ which is kept constant when the processes
creating or destroying $X$ are frozen.  The value of $Y$ varies from
$\sim10^{-2}$ to $10^{-9}$.  The number density of $X$ at temperature $T$ is
given by $n_X(T)=Ys(T)$, where $s(T)=\frac{2\pi^2}{45}g_{*s(T)}T^3$ is the
entropy density. For $T\ll m$, $X$ is nonrelativistic and the energy density
is given by $\rho_X(T)=mn_X(T)$.

Let us consider the relevant range of the values of $mY$ and $\tau$.
There are two obvious bounds on $mY$ and $\tau$.  The first concerns
nucleosynthesis.  At the time of neutrino decoupling, one customarily requires
that the energy density of $X$ should be less than that of one neutrino
species.  This imposes an upper bound on $mY$ which applies to the case that
$m>1$ MeV,
\begin{equation}
\mY < 0.107. \label{nucleo}
\end{equation}
The second comes from the condition that the energy density of the decay
products of $X$ should not overclose the universe:
\begin{equation}
\ts\mY^2 < 2\times10^6h^3. \label{closure}
\end{equation}
If $X$ can decay into photons or ordinary particles, a severe restriction can
be imposed by the present observation of microwave background radiation
\cite{eglns92}.  Therefore, we consider cases in which the branching ratio of
X into photons is negligible and decay products of X are very weakly
interacting so as not to disturb the spectra of observable particles.

In this scenario, there exist two eras of matter domination.  The first era
begins when the energy density of the late decaying particles dominates the
energy density of the universe.  The first matter dominated era ends when the
late decaying particle decays into extremely relativistic particles.  The
second era begins when the energy density of cold dark matter dominates over
the energy density of the decay products.  The evolution of the energy
densities is shown in Fig.~2.  Four interesting scale factors appear in this
scenario: $R_{EQ1}$ (the scale factor at which the energy density of $X$'s and
the energy density of photons and neutrinos become equal), $R_D$ (the scale
factor at which $X$ decays), $R_{EQ}$ (the scale factor at which the energy
density of cold dark matter and the energy density of photons and neutrinos
become equal), and $R_{EQ2}$ (the scale factor at which energy densities of
cold dark matter and extremely relativistic decay products become equal).
$R_{EQ}$ is the scale factor at the radiation--matter equality epoch in the
simple CDM model.  In our scenario, we make $R_{EQ2}>R_{EQ}$.

The temperature and the time at which the first matter domination begins are
given by
\begin{eqnarray}
T_{EQ1} &=& 1.55\ mY, \\
t_{EQ1} &=& 0.55\mYi^2{\rm\ sec},
\end{eqnarray}
if the lifetime $\tau$ is larger than $t_{EQ1}$.
Actually, the conditions for which $X$-domination really occurs are
$\tau>t_{EQ1}$ and $t_{EQ1}<t_{EQ}$, which are converted to
\begin{eqnarray}
 &\displaystyle \ts\mY^2 > 0.55, & \label{Xdomination1} \\
 &\displaystyle \mY > 3.6\times10^{-6}h^2. &
\label{Xdomination2}
\end{eqnarray}
To find out the scale factor $R_{EQ2}$, we use the rough approximation of
simultaneous decay and sudden change of radiation and matter domination. The
condition for $R_{EQ}$ and $R_{EQ2}$ read
\begin{eqnarray}
\rho_0\left(\frac{R_0}{R_{EQ}}\right)^3 &=&
\rho_{EQ1}\left(\frac{R_{EQ1}}{R_{EQ}}\right)^4 \\
\rho_0\left(\frac{R_0}{R_{EQ2}}\right)^3 &=&
\rho_{EQ1}\left[\left(\frac{R_{EQ1}}{R_D}\right)^3
\left(\frac{R_D}{R_{EQ2}}\right)^4
+\left(\frac{R_{EQ1}}{R_{EQ2}}\right)^4\right]
\end{eqnarray}
where $\rho_{EQ1}$ is the energy densities of $X$ at the first radiation matter
equality and $\rho_0$ is the present matter energy density.  From these, it
follows that the second matter domination in our scenario occurs later than the
well-known matter domination of the simple CDM model by a factor
\begin{equation}
\frac{R_{EQ2}}{R_{EQ}}
\simeq \frac{R_D}{R_{EQ1}}+1
\simeq \left(\frac{\tau}{t_{EQ1}}\right)^{2/3}+1,
\end{equation}
where the last equality comes from the relation $R\propto t^{2/3}$ during the
matter dominated era.

The evolution of the fluctuation spectrum is obtained by the linear
perturbation theory as done in Ref.~\cite{be91} for the 17 keV neutrino case.
But here, we present a rough behavior and a simple estimate.  The fluctuation
spectrum is characterized by two length scales, associated with the horizon
lengths at two instances of radiation matter equality,
\begin{eqnarray}
\lambda_{EQ1} & \simeq &
      8\times10^{-2}\mYi{\rm\ kpc}, \\
\lambda_{EQ2} & \simeq & 30(\Omega h^2)^{-1}
\left[\left\{\frac{1}{0.55}\ts\mY^2\right\}^{2/3}+1\right]^{1/2}{\rm\ Mpc}.
\label{horizon2}
\end{eqnarray}

The existence of the scale $\lambda_{EQ1}$ distinguishes this scenario from
the standard CDM model.  Objects on this and smaller scales would collapse at
high red shifts and more structures exist on these scales than the standard
CDM model.  We will discuss about it later.

$\lambda_{EQ2}$ is larger than $\lambda_{EQ}$ when the X-domination condition,
eq.~(\ref{Xdomination1}) holds.  In the previous section, we mentioned that
the $\Omega h=0.2$ CDM model have a good fit with possible small biasing taken
into account.  Lowering $\Omega h$ has an effect of making $\lambda_{EQ}$
large.  But this model is in conflict with the inflationary paradigm
$\Omega=1$ and the observed Hubble constant $h\simeq0.4-1$.  In the late
decaying particle scenario, we have the same effect of making $\lambda_{EQ}$
large but still can keep $\Omega=1$.  The value $\Omega h=0.2$ gives a rough
estimation of the desired value for $\lambda_{EQ2}$:
$\lambda_{EQ2}/h^{-1}{\rm\ Mpc}\simeq150$.  It can be rewritten as a condition
on $\tau$ and $mY$ for $\Omega=1$,
\begin{equation}
\ts\mY^2 \simeq 0.55 \left[\left(h/0.2\right)^2-1\right]^{3/2}.
\label{main_for}
\end{equation}
This together with the eqs.~(\ref{nucleo}) and (\ref{Xdomination2}) forms a
main condition of the late decaying particle scenario with $\Omega=1$.

\section{Models of A Late Decaying Particle}

The long lifetime and the small radiative branching ratio of a late decaying
particle make it difficult to construct particle physics models for it.
But it is certainly not impossible.  It is possible to use symmetry and/or
coupling suppressed by a large mass scale to obtain the long lifetime and to
suppress the radiative decay.  In this section, we present two models: the
massive neutrino and the light axino.

\subsection{A Late Decaying Neutrino}

Cosmological implications of the massive neutrino vary depending on the size
of mass and lifetime.  At present, the stable $\sim30{\rm\ eV}$ mass neutrino
is a good candidate for hot dark matter.  Here we consider a different region
in mass and lifetime.
The relic abundance of a light neutrino species is $Y=3.9\times10^{-2}$.
The late decaying particle scenario works for the neutrino mass in the range
$100{\rm\ eV} \ls m_\nu \ls 1{\rm\ MeV}$.
Hereafter, we use $m_\nu=17{\rm\ keV}$ as the representative value and adopt
the notation $m_{17}=m_\nu/17{\rm keV}$.  Then $mY=0.66m_{17}$ keV and the
eq.~(\ref{main_for}) demands the lifetime to be
\begin{equation}
\tau\simeq0.5m_{17}^{-2}{\rm\ yr},  \label{nlt}
\end{equation}
for the standard value $h=0.5$.
(Actually, the use of the formulae of the previous section needs a slight
modification due to the change in the number of relativistic neutrino
species.  But we still use them because the estimations are not affected
very much by this modification.)
The length scale of the first radiation matter equality is
\begin{equation}
\lambda_{EQ1}\simeq120m_{17}^{-1}{\rm\ kpc}.
\end{equation}

The well-known way to obtain the neutrino mass in the above mentioned range is
to use the seesaw mechanism.  The long lifetime required in the eq.~(\ref{nlt})
can be obtained in several models \cite{fy85}.  They considered schemes in
which the massive neutrino decays into exotic particles --- majoron, familon or
techniphoton.  In either case, we have to make a judicious choice of unknown
parameters to give such a long lifetime.

\subsection{A Light Axino and A Lighter Gravitino}

The importance of the cosmological effects of axinos and gravitinos arises from
the weaknesses of their interactions due to the suppression factors of the
axion decay constant for the axino and the Planck mass for the gravitino.
Here we focus on the scenario in which the gravitino is the lightest
supersymmetric particle (LSP) and the axino is the second lightest
supersymmetric particle (SLSP).

For this, we need the simultaneous implementation of local supersymmetry and
the axion solution of the strong $CP$ problem. Supersymmetry has been
introduced to solve the gauge hierarchy problem. Realistic models incorporating
low energy supersymmetry consider local supersymmetry (supergravity),
spontaneously broken by the hidden sector \cite{n84}.  Most of them have an $R$
symmetry to suppress the unwanted large $B$ and $L$ violations.  The $R$
symmetry dictates the existence of a stable particle commonly called the LSP
which is a good candidate for dark matter.  The LSP is model dependent and a
popular candidate is neutralino, a linear combination of neutral gauginos and
neutral higgsinos.  If we adopt local supersymmetry, the gravitino is another
natural candidate for the LSP.  Peccei-Quinn symmetry is needed for a solution
of the strong $CP$ problem, and no idea seems to be compelling in this regard
\cite{k87}.  It predicts the existence of the axion \cite{pq77} and the
coherent classical axion oscillation created at $T\simeq\Lambda_{\rm QCD}$
during the expansion of the universe is a good candidate for dark matter
\cite{pww83}.  The simultaneous implementation of these two ideas can affect
the status of cold dark matter candidates \cite{rtw91}.  The axino (the
superpartner of axion) can and generally do destabilize the lightest ordinary
supersymmetric particle (LOSP) which is a popular dark matter candidate.  Then
the viable candidates for dark matter are gravitinos, axinos and axions.  For
the gravitino and the axino to be LSP and SLSP, we need gravitino and axino
lighter than the LOSP.

Currently favored gravitino mass is identified with the electroweak scale of
$10^2-10^3$ GeV and the gravitino with such mass cannot be the LSP because
there is usually a lighter supersymmetric particle to which the gravitino can
decay.  Recently, however, the old idea of supercolor at multi TeV region has
been reinvestigated without phenomenological difficulties \cite{dn93}.  In this
case the gravitino mass falls in the $0.4$ meV $-$ 0.4 keV region for
$\Lambda_{\rm supercolor}=1-10^3$ TeV.  Therefore, the phenomenologically
favored gravitino mass is still open in these two regions and it is worthwhile
to consider the cosmological consequence of this light gravitino scenario.

In the case that $R$ symmetry is exact and the axino is lighter than the LOSP,
the LSP lighter than the axino must exist for the axino to decay.  The
presently known candidate is the gravitino.  The possibility and cosmological
constraints of the gravitino as the LSP were previously considered \cite{b91}.
Here, we summarize the results.  Gravitinos typically decouple just after the
Planck time.  From this one obtains a gravitino mass bound $m_{3/2}\le 2$ keV
in the standard big bang cosmology.  But the inflation after the gravitino
decoupling might have washed  out relic gravitinos almost completely and we
have no such bound. Gravitinos can be regenerated through the reheating
process subsequent to inflation.  The requirement that the regenerated
gravitinos should not overclose the universe gives a constraint between the
reheating temperature and the gravitino mass \cite{ekn84}.
If we assume that the reheating temperature is lager than the axino decoupling
temperature $10^{11}$ GeV, a region of gravitino mass from a few keV to a few
$\times$ 10 GeV  is excluded.  There is also a bound in the extremely light
gravitino.  For the very light gravitino whose mass $m_{3/2}$ is small compared
to the typical interaction energy, the helicity $\pm\frac{1}{2}$ component of
the gravitino dominate the interaction and the strength of the interaction
becomes larger as $m_{3/2}$ becomes smaller.  Therefore, the extremely light
gravitino ($m_{3/2}\ll1$ keV) interacts strongly and decouples well below the
weak scale.  The standard nucleosynthesis scenario constrains the number of the
species of the neutrino-like particles to be smaller than 3.3 at the time of
nucleosynthesis.  Then, gravitinos must decouple before $T\simeq 200$ MeV to
get a sufficient dilution factor. This yields a gravitino mass bound
\begin{equation}
m_{3/2} \gs 10^{-4}\left(\frac{m_{\tilde l}}{100{\rm GeV}}\right){\rm\ eV}
\end{equation}
where $m_{\tilde l}$ is the slepton mass.

The axino mass is predicted to be in a region between 10 GeV and keV in
supergravity models with supersymmetry broken in the hidden sector.  Usually
one adopts the minimal kinetic energy term.  In general, there exist other loop
contributions and the kinetic energy term can be of nonminimal form.
Furthermore, we can adopt the supersymmetry breaking by supercolor type
interactions. Therefore, the axino mass heavier than keV is a theoretical
possibility.

Now, we calculate the axino lifetime. The dominant coupling of axion to
gravitino is
\begin{equation}
{\cal L}_{a\tilde a\tilde G}=
{1\over M}\bar\psi_\mu \gamma^\nu\partial_\nu z^{*}\gamma^\mu \tilde a_L
+{\rm h.c.}
\end{equation}
where $z=(s+ia)/\sqrt{2}$ in terms of saxino $s$ and axion $a$.  There exists
another interaction term proportional to $m_{3/2}$ which is negligible
compared to the above interaction.  In the case that the axino is heavier than
the gravitino, the axino decays into the gravitino and the axion.  In the
limit of vanishing axion mass, the axino lifetime is
\begin{equation}
\tau_{\tilde a} = {96\pi M^2 m_{3/2}^2\over m_{\tilde a}^5}
= 1.2\times 10^{12}\left({{\rm MeV}\over m_{\tilde a}}\right)^5
\left(m_{3/2}\over {\rm eV}\right)^2 {\rm\ sec.}
\end{equation}

Axinos are kept in thermal equilibrium at high temperature through the process
$q\bar q\leftrightarrow\tilde a\tilde g$ with an intermediate gluon.
{}From the interaction lagrangian
\begin{equation}
{\cal L}=\frac{\alpha_c}{8\pi F_a}\left(aG^a_{\mu\nu}\widetilde G^{a\mu\nu}
+sG^a_{\mu\nu}G^{a\mu\nu} +
\overline{\tilde a}\gamma_5\sigma_{\mu\nu}\tilde g^aG^{a\mu\nu}
+\cdots\right),
\end{equation}
the reaction rate can be calculated to give
$\Gamma\sim\alpha_c^3T^3/16\pi F_a^2$ for $T<F_a$.  On the other hand, the
expansion rate is $H=1.66g_*^{1/2}T^2/M_{pl}$ where $g_*=915/4$ for the
particle content in the minimal supersymmetric standard model.  Therefore, the
decoupling temperature is
\begin{equation}
T_{\tilde a} = 10^{11}\left(\frac{F_a}{10^{12}{\rm GeV}}\right)^2
\left(\frac{0.1}{\alpha_c}\right)^3 {\rm\ GeV}.
\end{equation}

The amount of relic axinos depend on the decoupling temperature.
(There are also axinos from decays of relic LOSPs. But the number density of
relic LOSPs is usually much smaller than that of hot thermal relic and hence
axinos from decays of relic LOSPs can be neglected.)
The axino decoupling temperature $T_{\tilde a}$ is so high that it can be
comparable to or larger than the reheating temperature $T_R$. For simplicity,
we consider two extreme cases which differ from each other significantly.

({\it i\/}) $T_R\gg T_{\tilde a}$:
The relic axinos are hot thermal relic so that $Y\simeq1.8\times10^{-3}$.
The desired value of $\lambda_{EQ2}$ is obtained for
\begin{equation}
\left(\frac{{\rm MeV}}{m_{\tilde a}}\right)^3
\left(\frac{m_{3/2}}{{\rm eV}}\right)^2
\simeq 1.4\times10^{-7}(25h^2-1)^{3/2},  \label{region1}
\end{equation}
within the mass range
\begin{eqnarray}
 & 2{\rm\ keV} \ls m_{\tilde a} \ls 60{\rm\ MeV}, &  \label{region2} \\
 & \displaystyle \left(\frac{{\rm MeV}}{m_{\tilde a}}\right)^3
\left(\frac{m_{3/2}}{{\rm eV}}\right)^2 < 0.5h^3, &  \label{region3}
\end{eqnarray}
which come from the eqs.~(\ref{nucleo}), (\ref{closure}) and
(\ref{Xdomination2}).  This requires  the MeV mass axino and the very light,
sub-eV mass gravitino.  The cosmologically allowed mass ranges and the region
corresponding to the eq.~(\ref{region1}) are shown in Fig.~3.  The smaller
characteristic length scale is given by
\begin{equation}
\lambda_{EQ1} = 44 \left(\frac{{\rm MeV}}{m_{\tilde a}}\right){\rm\ kpc}.
\label{scls1}
\end{equation}
With the MeV mass axino, it is about the globular cluster scale.

({\it ii\/}) $T_R\ll T_{\tilde a}$:
The hot thermal relic axinos are diluted away by inflation and relic axinos
are those regenerated through the nonequilibrium processes, which yields
$Y\simeq7\times10^{-6}\widetilde{F}_a^{-2}\widetilde{T}_R$
where $\widetilde{F}_a=F_a/10^{12}{\rm GeV}$ and
$\widetilde{T}_R=T_R/10^6{\rm GeV}$.
Now the corresponding equations of the eqs.~(\ref{region1})--(\ref{scls1})
contain $F_a$, $T_R$ as well as $m_{\tilde a}$, $m_{3/2}$ and are given by
\begin{equation}
\widetilde{F}_a^{-4} \widetilde{T}_R^2
\left(\frac{{\rm GeV}}{m_{\tilde a}}\right)^3
\left(\frac{m_{3/2}}{{\rm keV}}\right)^2
\simeq 8.8\left(25h^2-1\right)^{3/2},
\end{equation}
\begin{eqnarray}
 & \displaystyle
5\times10^{-4} \ls
\widetilde{F}_a^{-2}\widetilde{T}_R
\left(\frac{m_{\tilde a}}{{\rm GeV}}\right) \ls 15, & \\
 & \displaystyle
\widetilde{F}_a^{-4}\widetilde{T}_R^2
\left(\frac{{\rm GeV}}{m_{\tilde a}}\right)^3
\left(\frac{m_{3/2}}{{\rm keV}}\right)^2 < 3\times10^7h^3, &
\end{eqnarray}
and
\begin{equation}
\lambda_{EQ1} = 11
\widetilde{F}_a^2 \widetilde{T}_R^{-1}
\left(\frac{{\rm GeV}}{m_{\tilde a}}\right)^2 {\rm\ kpc}.
\end{equation}
This case opens a possibility of the heavier axino with GeV mass depending on
the size of the reheating temperature.

\section{Discussion and Conclusion}

In our scenario of late decaying particles, there exists a brief era of matter
domination between $\lambda_{EQ1}$ and $\lambda_D$.  During this epoch, there
exists a possibility that some scales can grow and form structures.  Because
there exists a characteristic scale of globular clusters, it is worthwhile to
see if the structure formation is triggered during this brief period and to
check whether it coincides with the scale of globular clusters.  If the
conclusion is affirmative, then the scenario of the late decaying particle
provides an explanation for the size of globular clusters.
Furthermore, such a large power at scale smaller than the galaxy scale is
advantageous on explaining the Lyman-line-absorbing clouds seen in quasar
spectra \cite{abc85}.  The problem is that it leads to large anisotropies in
microwave background radiation at small angles.  But it is known that this
early structure formation, e.g. the early population of quasi-stellar objects,
could reionize the universe shortly after the normal epoch of photon--baryon
decoupling at $z\sim1000$, thereby reducing small angle anisotropies in the
microwave background radiation.

There are a few merits of the axino-gravitino scenario.  First, it has the
classical axion oscillation as a natural candidate for cold dark matter.
Second, the interactions of the axino and the gravitino are well known so that
the unknown parameters are mainly their masses and no unknown coupling is
involved.  Third, the explanation of the globular cluster scale or other
sub-galaxy structures may fix their masses.

In summary, we considered the late decaying particle scenario in the $\Omega=1$
CDM model as a viable alternative to the simple CDM model which failed in
fitting all the observational data simultaneously.  The late decaying
particles delay the beginning of the matter domination, thereby reducing the
excessive power at small scales which was the problem of the simple CDM model.
This scenario has another peak in the power spectrum developed during the
earlier matter domination.  Its length scale is smaller than the galaxy scale,
thereby having implications for sub-galaxy structures, which need further
investigations.  The particle physics model realizing this scenario is
difficult to construct due to the required long lifetime and small radiative
branching ratio of a late decaying particle, but certainly not impossible.  We
presented two examples.  The keV mass neutrino with lifetime of order 1 yr,
which has been considered by many authors, is a possible candidate of the late
decaying particle.  The axino-gravitino scenario requires the MeV mass axino
and the sub-eV mass gravitino (the sufficiently low reheating temperature can
raise the required masses).  It possesses all the necessary ingredients of the
late decaying particle scenario with the simultaneous implementation of local
supersymmetry and Peccei-Quinn symmetry.

\acknowledgments

We have benefitted from discussions with E. J. Chun who participated in this
research at the early stage of collaboration.
We thank C. Park for helpful discussion.
This work is supported in part
by the Korea Science and Engineering Foundation through Center for Theoretical
Physics at Seoul National University, KOSEF-DFG Collaboration Program, and the
Ministry of Education through the Basic Science Research Institute, Contract
No.~BSRI-94-2418.

\begin{figure}
\caption{Comparison of the CDM model power spectrum with the observational
data.  The solid line corresponds to the $\Omega h=0.3$ CDM model, while the
dashed and dotted lines correspond to the $\Omega h=0.5$ and $1$ CDM models
respectively.}
\end{figure}

\begin{figure}
\caption{The evolution of energy densities of radiation $(\gamma,\nu)$, late
decaying particles $(X)$, their decay products and the cold dark matter.}
\end{figure}

\begin{figure}
\caption{Cosmologically allowed region in the axino mass -- gravitino
mass plane.  Dotted regions are excluded by the closure bound and the
nucleosynthesis bound (the equation numbers represent the relevant equations
in the text).  The heavy dotted line corresponds to the
eq.~(\protect\ref{region1}) for $h=0.5$.  The white island below the dashed
line is the mass range allowed by the eqs.~(\protect\ref{region2}) and
(\protect\ref{region3}).}
\end{figure}

\end{document}